\documentclass[aps,prb,floats,twocolumn,floatfix]{revtex4}
\usepackage{psfig,epsf,graphicx}
\usepackage{amssymb}
\usepackage{eepic}

\newlength{\piclen}
\renewcommand{\Re}{Re }
\renewcommand{\Im}{Im }

\begin{document}

\title{Electronic Structure of Paramagnetic $\rm {\bf V_2O_3}$:\\
Strongly Correlated Metallic and Mott Insulating Phase}

\author{G.\ Keller$^1$, K.\ Held,$^2$ V.\ Eyert$^1$, D.\ Vollhardt,$^1$ and V.\ Anisimov$^3$}
\affiliation{$^1$Institut f\"ur Physik, Universit\"at Augsburg,
         86135 Augsburg, Germany\\ $^2$Max-Planck-Institut f\"ur
Festk\"orperforschung, 70569 Stuttgart, Germany\\
$^3$Institute of Metal Physics, Ekaterinburg GSP-170, Russia }

 \date{Version 1, \today}

\begin{abstract}
{\large
LDA+DMFT, the computation scheme merging the local density approximation
and the dynamical mean-field theory, is employed to calculate spectra
both below and above the Fermi energy and spin and orbital occupations
in the correlated paramagnetic metallic and
Mott insulating phase of V$_{2}$O$_{3}$. The self-consistent DMFT equations
are solved by quantum Monte Carlo simulations.
Room temperature calculations provide direct comparison with experiment.
They show a significant increase of the quasiparticle height in comparison
with the results at 1160\ K. We also obtain new insights into the nature of
the Mott-Hubbard transition in  V$_{2}$O$_{3}$. Namely, it is found to be
strikingly different from  that in the one-band Hubbard model due to the
orbital degrees of freedom.
Furthermore we resolve the puzzle of the unexpectedly small Mott gap in Cr-doped
 V$_{2}$O$_{3}$.}
\end{abstract}

\maketitle

\section{Introduction}
\label{sect1}

\begin{figure}[htb]
{\unitlength0.35cm
\includegraphics[clip=true,width=\piclen]{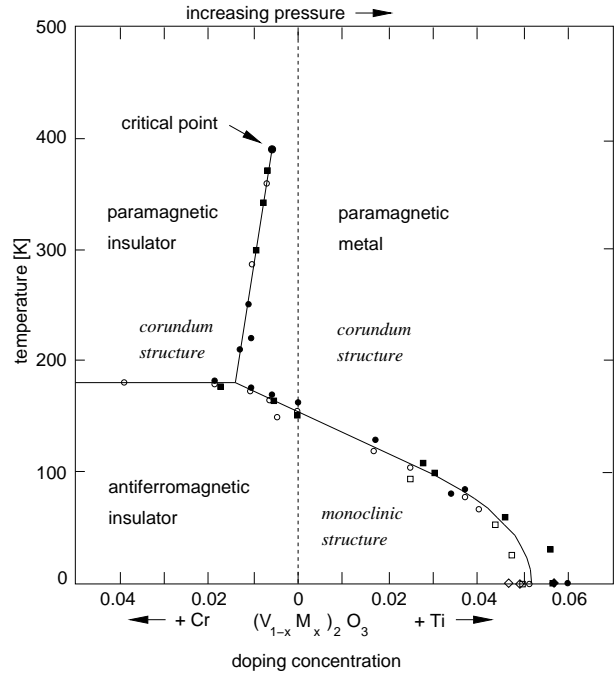}\\
\caption{Phase diagram of V$_{2}$O$_{3}$ showing the MIT as a function
of pressure and of doping with Cr and Ti; data
points from McWhan {\em et al.}~\cite{mcwhan73b}}
\label{fig:phase}}
\end{figure}

The phase transition between a paramagnetic metal and a paramagnetic
insulator caused by the Coulomb interaction between the
electrons is referred to as Mott-Hubbard metal-insulator
transition.~\cite{Mott,Gebhard} Reliable microscopic investigations of this many-body
phenomenon are known to be exceedingly difficult. Indeed, the
question concerning the nature of this transition poses one of the
fundamental theoretical problems in condensed matter physics.
Correlation-induced metal-insulator transitions
(MIT) of this type are found, for example, in transition metal
oxides with partially filled bands near the Fermi level. In these
systems band theory typically predicts metallic
behavior. The most famous example is V$_{2}$O$_{3}$ doped with Cr;~\cite%
{mcwhan70,mcwhan73b,RMcWh} see Fig. \ref{fig:phase}. While at low temperatures
V$_{2}$O$_{3}$ is an antiferromagnetic insulator (AFI) with
monoclinic crystal symmetry, the high-temperature paramagnetic
phase has a corundum structure. All transitions shown in the phase diagram are of
first order. In the case of the transitions from the
high-temperature, paramagnetic phases into the low-temperature
antiferromagnetic phase this is naturally explained by the fact
that the transition is accompanied by a change in crystal
symmetry. By contrast,  the MIT in the
paramagnetic phase is iso-structural;  only the ratio of the
$c/a$ axes changes discontinuously. This may be taken as an indication
for a predominantly electronic origin of this transition.

To explain an MIT induced by electronic correlations one can either
investigate a simplified electronic many-body model
to understand, at least, some of the basic features of the MIT, or
employ material-specific approaches such as the density functional
theory in the local density approximation (LDA). Concerning the former
approach,~\cite{RMcWh,Mott} the spin $S=1/2$, half-filled,
single-band Hubbard model~\cite{HubbardI,Gutzwiller,Kanamori} is
certainly the simplest possible model to be investigated. In
particular, the existence of an MIT in the paramagnetic phase
 of the half-filled Hubbard model had been
investigated already in the early work of Hubbard.~\cite{HubbardI,HubbardIII}
However, while the Hubbard I and III approximations~\cite{HubbardI,HubbardIII}
describe the insulating phase rather well, they do not describe a Fermi liquid
phase on the metallic side. On the other hand, the Gutzwiller approximation
provides a picture of the break-down of the Fermi liquid phase as
indicated by the collapse of the quasiparticle peak and the simultaneous
divergence of the effective mass at a critical value of the Coulomb interaction  $U_c$
(Brinkman-Rice transition).~\cite{BR}  However, within this framework one cannot describe the
Hubbard bands which are essential both in the strongly correlated
metallic phase below $U_c$ and in the insulating phase above $U_c$.
With these limitations, the details of the MIT in the Hubbard model remained unclear,
except for the one-dimensional case~\cite{Lie68} which is very particular
since it always describes an insulator, i.e., $U_c=0^{+}$.

During the last few years, our understanding of the MIT in the
one-band Hubbard model has considerably improved due to the
development of dynamical mean-field theory (DMFT).~\cite{DMFT_vollha, DMFT_georges} Within DMFT the
electronic lattice problem is mapped onto a self-consistent
single-impurity Anderson model.~\cite{DMFT_georges} This mapping becomes exact in the
limit of infinite coordination number~\cite{DMFT_vollha} and allows one to
investigate the dynamics of correlated lattice electrons
non-perturbatively at all interaction strengths. This is of
essential importance for a problem like the MIT which occurs at a
Coulomb interaction comparable to the electronic band-width.
In particular, DMFT provides a framework for deriving a coherent
picture of the electronic spectrum at all energy scales, i.e., of
the incoherent features at high energies (Hubbard bands),~\cite{HubbardIII}
 and the coherent quasiparticle behavior at low
energies.~\cite{Gutzwiller, brinkman70} At $T=0$, the
transition from the metallic to the insulating state is
signaled by a divergence of the effective mass and 
the collapse of the Fermi liquid quasiparticle peak
at the Fermi energy for Coulomb interaction  $U\rightarrow U_c^{-}$.~\cite{DMFT_georges,DMFTMott_a,DMFTMott_b,DMFTMott_c}
DMFT furthermore revealed the coexistence of
the metallic and the insulating phase below
a critical point at temperature $T_c$, such that there is a first order
phase transition in agreement with the experimental
observation for V$_{2}$O$_{3}$.
To investigate the MIT in ${\rm V_{2}O_{3}}$,
Rozenberg {\em et al.}~\cite{Rozenberg95} applied DMFT to
the one-band Hubbard model.
The influence of orbital degeneracy was studied by means of the
two-~\cite{Rozenberg97a,Han98a,Held98a} and three-band~\cite{Han98a}
Hubbard model for the semicircular density of states (DOS) of a Bethe lattice.
Most recently, a detailed analysis~\cite{Limelette03} of the conductivity change
demonstrated that, except for a very narrow region directly
at the critical point, the critical exponents are
of the liquid-gas transition type, in accordance with
a Landau theory for the Mott transition within DMFT.~\cite{DMFTMott_d,DMFTMott_e,DMFTMott_a}

Although the Hubbard model is able to explain certain basic
features of the Mott-Hubbard MIT in V$_{2}$O$_{3}$ and its phase
diagram, it cannot explain the physics of that material in any
detail. Clearly, a realistic theory of ${\rm V_{2}O_{3}}$ must
take into account the complicated electronic structure of this system.
In our previous work,~\cite{held01prl} we therefore applied the LDA+DMFT scheme to study the
MIT in paramagnetic V$_2$O$_3$.~\cite{Anisimov97,Held01} With LDA spectra
calculated for the crystal structure of metallic V$_2$O$_3$ and insulating $ {\rm (V_{0.962}Cr_{0.038})_2O_3} $ as input for the
subsequent three-band DMFT(QMC) calculations, we found an MIT, or rather a sharp crossover, at $U\approx 5$ eV.
Due to  restrictions in computer resources, the QMC calculations in Ref.~\onlinecite{held01prl} were
done at $T=1160$ K.
Subsequently, extensive QMC simulations at temperatures down to $T\approx 300$ K
were performed to make possible a comparison between experiment and theory at experimentally
relevant temperatures. Those computations yielded
spectra with a quasiparticle peak at the Fermi edge considerably stronger than that at $1160$~K and
were in contrast to the existing photoemission measurements. This puzzle
was finally resolved by recent improvements in photoemission spectroscopy (PES) experiments
which allowed one to perform high-energy, bulk sensitive PES, displaying a
prominent peak at $E_F$ in essential agreement with the LDA+DMFT results.~\cite{mo02}

In this paper, we provide details of our calclations reported in Ref.~\onlinecite{mo02}, present
LDA+DMFT spectra for $300$~K, $700$~K and $1160$~K both below and above the Fermi edge, and
compare them to PES and XAS measurements. Based on calculations of spin and orbital occupations, we then discuss the
properties of the ground state. In particular, the
nature of the MIT turns out to be rather distinct from that in a one-band model, i.e., we find that the
effective mass in the $a_{1g}$ orbitals does {\em not} diverge at the MIT transition.

The paper is organized as follows: In Section \ref{lda} the LDA band structure, based on the
experimental corundum crystal structure, and the
relevance of V-V pairs are discussed.
Electronic correlations are taken into account by DMFT as described in Section \ref{dmft}.
The resulting LDA+DMFT spectra are presented in  Section \ref{spectra},
including a discussion of the dependence on temperature
and the Hund's rule exchange coupling.
The pecularities of the MIT in V$_2$O$_3$ and the differences
to the MIT of the one-band Hubbard model are worked out in Section \ref{mit}.
 A detailed comparison with the experimental spectra
follows in Section \ref{comparison}. A summary and outlook is finally
provided in Section \ref{summary}.

\section{Crystal and electronic structure}
\label{lda}\label{sec3}

\begin{figure}[htb]
{\unitlength0.35cm
\begin{picture}(8,7)(0,5)
\put(0,0.5){\includegraphics[clip=true,width=3.5cm]{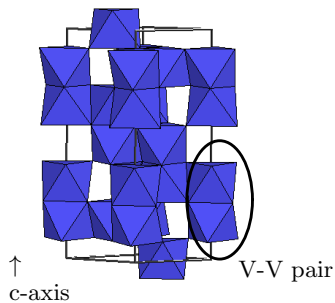}}
\put(8.0,4.4){\Thicklines \ellipse{2.5}{4.5}}
\put(8.4,1.3){\makebox(0,0)[bl]{ V-V pair}}
\put(0,1.5){\makebox(0,0)[bl]{$\uparrow$}}
\put(0,0.6){\makebox(0,0)[bl]{c-axis}}
\end{picture}}
\vspace{1.5cm}

\caption{Crystal structure of $\rm V_2O_3$.}
\label{fig:octahedra}
\end{figure}

In the paramagnetic metallic phase stoichiometric ${\rm V_2O_3}$
crystallizes in the corundum structure, which has a trigonal lattice and
space group $ R\bar{3}c $ ($ D_{3d}^{6} $) with lattice constants
$a  = 4.9515${\AA} and $c = 14.003${\AA}.~\cite{dernier70a} Vanadium
and oxygen atoms occupy the Wyckoff positions (12c) and (18e) with internal
parameters $ z_V = 0.34630 $ and $ x_O = 0.31164 $, respectively, which
deviate markedly from the value 1/3 assumed in an ideal hexagonal arrangement.~\cite{dernier70a}
Within the corundum structure the vanadium atoms are arranged in pairs
along the hexagonal $c$-axis, which can be derived from an ideal chain
structure by introducing vacancies at every third site.~\cite{mattheiss94}
The oxygen atoms form distorted octahedra around the vanadium sites.
While the V-V pair along the hexagonal $c$-axis
shares octahedral faces, the octahedra are interlinked via
edges and corners perpendicular to this axis,~\cite{dernier70a,mattheiss94} see Fig.~\ref{fig:octahedra}.

In the Cr-doped paramagnetic insulating phase
 the lattice symmetry is preserved, but the
crystal structure parameters change slightly. In particular, for
$ {\rm (V_{0.962}Cr_{0.038})_2O_3} $ the lattice constants amount to
$ a = 4.9985${\AA} and $ c = 13.912${\AA} and the positional parameters
are $ z_V = 0.34870 $ and $ x_O = 0.30745 $, respectively.~\cite{dernier70a}
All these changes combine into a distinct displacement pattern: As compared
to the metallic phase the shared octahedral faces
between the V-V pair shrink, while those octahedral faces
pointing to the opposite side along the $c$-axis, i.e.,
towards the aforementioned vacancies, are enlarged.  At the same time, the
vanadium atoms shift away from the shared faces of the V-V pair such that the
distances within the  V-V pair increase upon Cr-doping, even though the
$c$-axis lattice constant decreases. The increased  $a$-axis lattice constant
directly leads to enhanced vanadium distances within the $ab$-plane.
As a net result, {\em all} nearest-neighbor vanadium distances
are enlarged for insulating ${\rm (V_{0.962}Cr_{0.038})_2O_3}$.
Hence,  we expect a
reduction of the bandwidth especially of the $ t_{2g} $-derived bands.
From a comparison of pure and doped $ {\rm V_2O_3} $ as well as
$ {\rm Cr_2O_3} $, Dernier~\cite{dernier70a} concluded that the metallic properties are
intimately connected with the vanadium hopping within the $ab$-plane rather than with
hopping processes between the V-V pairs along the $z$-axis.

\begin{figure}[htb]
\includegraphics[clip=true,width=\piclen]{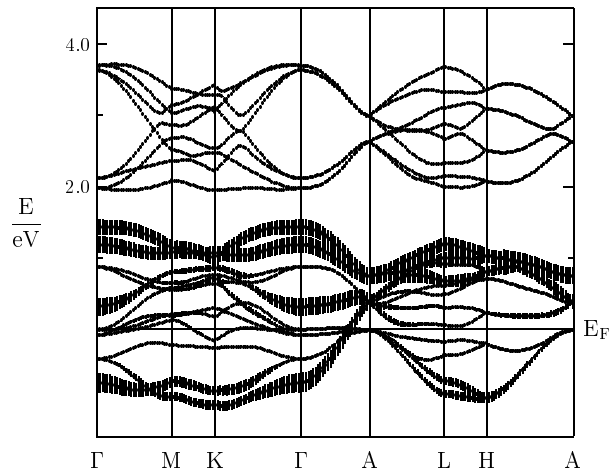}
\caption{Electronic bands of $ {\rm V_2O_3} $ along selected symmetry
         lines within the first Brillouin zone of the hexagonal
         lattice, Fig.\ \protect\ref{fig:bzhex}. The width of the bars
         given for each band indicates the contribution from the
         $ a_{1g} $ orbitals.}
\label{fig:bnd_met}
\end{figure}

\begin{figure}[htb]
\includegraphics[clip=true,width=\piclen]{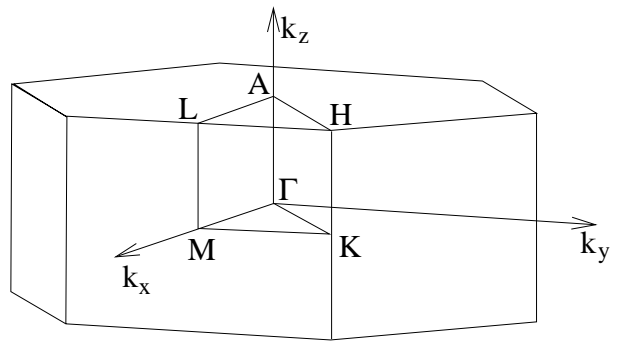}
\vspace*{5pt}
\caption{First Brillouin zone of the hexagonal lattice.}
\label{fig:bzhex}
\end{figure}

\begin{figure}[htb]
\includegraphics[clip=true,width=\piclen]{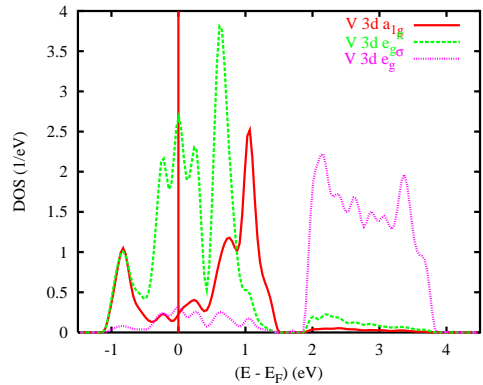}
\caption{Total and partial densities of states (DOS) of $ {\rm V_2O_3} $
         per unit cell.}
\label{fig:dos_met}
\end{figure}

\begin{figure}[htb]
\includegraphics[clip=true,width=\piclen]{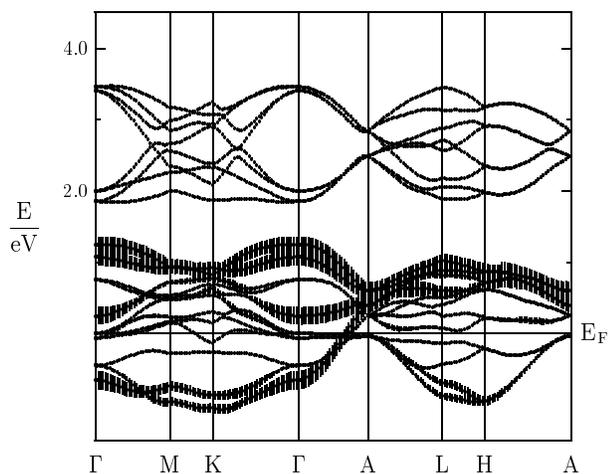}
\caption{Electronic bands of $ {\rm (V_{0.962}Cr_{0.038})_2O_3} $.}
\label{fig:bnd_iso}
\end{figure}

\begin{figure}[htb]
\includegraphics[clip=true,width=\piclen]{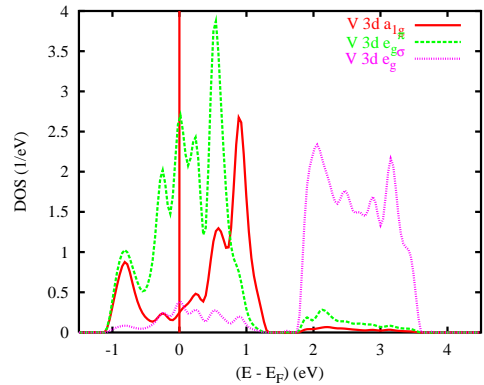}
\caption{Total and partial densities of states (DOS) of
         $ {\rm (V_{0.962}Cr_{0.038})_2O_3} $ per unit cell.}
\label{fig:dos_iso}
\end{figure}

In a first step LDA band structure calculations~\cite{hohenberg,kohnsham} were performed,
which used the augmented spherical wave (ASW) method.~\cite{wkg,revasw}
Figs. \ref{fig:bnd_met},\ref{fig:dos_met},\ref{fig:bnd_iso}, and \ref{fig:dos_iso} show these
bandstructures along selected high symmetry lines (Fig.\ \ref{fig:bzhex})
within the first Brillouin zone of the hexagonal lattice
and the densities of state (DOS) for  $ {\rm V_2O_3} $
and $ {\rm (V_{0.962}Cr_{0.038})_2O_3} $, respectively.

In total our results are in good agreement with those published by
Mattheiss.~\cite{mattheiss94} In particular, while the O $ 2p $ derived
bands show up in the range between -9 and -4 eV, the V $ 3d $ dominated
states fall, due to the octahedral surrounding with oxygen,
 into two groups of bands: $ t_{2g} $ and $e_g^{\sigma}$.
With this separation,  the  $e_g^{\sigma}$ bands will be empty
and the $t_{2g}$ bands partially filled with two electrons per V ion.

\begin{figure}[htb]
{\unitlength0.35cm
\includegraphics[clip=true,width=\piclen]{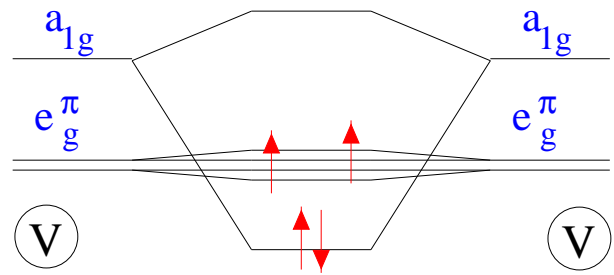}\\
\caption{Left and right: splitting of the $t_{1g}$ orbitals in the corundum
crystal structure. Middle: Formation of a chemical bond for a single V-V pair along the $c$-axis.}
\label{fig:pairing}}
\end{figure}

Due to the lower trigonal lattice symmetry the $t_{2g}$ states are further
split into doubly and singly degenerate $e_g^{\pi}$ and $a_{1g}$ states, see
 Figs.\ \ref{fig:dos_met}, \ref{fig:dos_iso} and the scheme Fig.\ \ref{fig:pairing}.
The value of this splitting
($\approx 0.3$\,eV for the centers of gravity) is much smaller than the $t_{2g}$ bandwidth
($\approx 2$\,eV). However, as the value of the Coulomb interaction
parameter $U$ ($U > 4$\,eV) is larger than the bandwidth, this small
trigonal splitting strongly determines the orbital ground state of the V-ion
obtained from LDA+DMFT calculations,
as will be shown below.
To highlight the difference between  $e_g^{\pi}$ and $a_{1g}$ states, we append
to each band at each $ {\bf k} $ point a bar in Fig.\ \ref{fig:bnd_met}, whose length is a measure for
the contribution from the $ a_{1g} $ orbitals to the respective wave function.

The changes on going to $ {\rm (V_{0.962}Cr_{0.038})_2O_3} $ are stated  easily: In  Figs.\ \ref{fig:bnd_iso}
and \ref{fig:dos_iso},
we observe a narrowing of the $ t_{2g} $ and $ e_g^{\sigma}$ bands of
$ \approx 0.2 $ and 0.1\,eV, respectively, as well as a slight downshift
of the centers of gravity of the $e_g^{\pi}$ bands. However, the insulating band gap
expected for a calculation with the insulating crystal structure is missing.
%However, the experimentally
%observed\cite{iso_gap} insulating band gap is missing.

As already mentioned a peculiarity of the corundum crystal
structure are the $c$-axis V-V pairs. Long ago Allen~\cite{allen} emphasized
the importance of the intra-pair
interactions for interpreting spectroscopic properties of V$_2$O$_3$
and its solid solution with Cr$_2$O$_3$.
Since the $a_{1g}$  orbitals
are directed along the c-axis, these orbitals are the ones which mediate a strong hybridization
between V-V pairs. This hybridization for the V-V pair led
Castellani {\em et al.}~\cite{castellani78} to
a model (see Fig.\ \ref{fig:pairing}) where two of the four
electrons per V-V pair occupy a bonding molecular orbital formed by
$a_{1g}$ orbitals, leaving two electrons (one per site) in a partially filled
twofold-degenerate $e_g^\pi$ band. That results in a spin-$\frac{1}{2}$ orbitally degenerate
state per V ion with complicated orbital and spin ordering pattern
explaining the unusual properties of the low-temperature antiferromagnetic phase.

Indeed, Figs.\ \ref{fig:bnd_met} and  \ref{fig:dos_met} show some splitting of
the $a_{1g}$  bands, in particular,  between the M and the K point of the bandstructure,
as one would expect from the formation of a chemical bond.
But, the situation is far more complicated than a simple chemical bonding of the $a_{1g}$ band into
a bonding and an anti-bonding band:
There is some additional spectral weight near $E_F$ (e.g.,  in the vicinity of the $\Gamma$ point).
Moreover,  there is not even a low-lying ``bonding'' $a_{1g}$ band in parts of the Brillouin zone
(e.g. between the H and A point).

The Castellani {\em et al.} model~\cite{castellani78} was challenged
by Park {\em et al.}~\cite{park} Based on the polarization dependence
of  x-ray absorption experiments they came to the conclusion that the
V$^{3+}$ ion is in a spin-1 state. They also demonstrated that
the orbital ground state of the ion is predominantly $e_g^\pi e_g^\pi$ with a small
admixture of $e_g^\pi a_{1g}$ configurations. This was later
supported by  LDA+U calculations of Ezhov {\em et al.}~\cite{ezhov99}
where a spin-1 ground state with a $e_g^\pi e_g^\pi$ orbital configuration
was obtained.

Nevertheless, the picture where the strongest hybridization parameter
in V$_2$O$_3$ is the $a_{1g}-a_{1g}$ hopping within the V-V pair, with all other
hybridizations being much smaller, is still popular. Many theoretical studies of
this material start with a (as good as possible) solution  for the V-V pair
and consider the inter-pair interactions as a perturbation.~\cite{mila,tanaka,mateo}
All these model calculations were based on the values of hopping parameters
obtained by least-square fit of LDA bands to a model Hamiltonian with nearest-neighbor hopping.
Recently this problem was reexamined by Elfimov {\em et al.}~\cite{elfimov} who found that the value
of the $a_{1g}-a_{1g}$ hopping in the V-V pair is significantly reduced if next-nearest
neighbor hoppings (which were found to be significant) are taken into account
in the fit to a model Hamiltonian. Hence, one cannot consider inter-pair
hoppings as a mere perturbation as it was taken for granted for a long time.

\section{Including electronic correlations via DMFT}
\label{sect4}\label{dmft}
The LDA band structure of the previous section clearly fails to
describe (V$_{1-x}$Cr$_x$)$_2$O$_3$. In particular, the chromium doped compound
($ {\rm (V_{0.962}Cr_{0.038})_2O_3} $ is an insulator whereas
LDA predicts metallic behavior.
The reason for this failure is that LDA deals with electronic correlations
only very rudimentarily, namely,
the dependence of the LDA exchange-correlation energy on the
electron density is given by
 perturbative or quantum  Monte-Carlo calculations
for jellium,~\cite{jellium,jellium2} which is a weakly correlated system.
To overcome this shortcoming, we supplement the LDA band structure
by the the most important Coulomb interaction terms, i.e., the
local Coulomb repulsion $U$ and the local Hund's rule exchange $J$.
The local Coulomb repulsion $U$ gives rise to a genuine
effect of electronic correlations,  the
Mott-Hubbard metal insulator transition.~\cite{DMFT_georges,DMFTMott_a,DMFTMott_b,DMFTMott_c,DMFTMott_d,DMFTMott_e,bulla_costi}
If the LDA bandwidth is considerably larger than the local
Coulomb interaction, the LDA results are slightly modified but
the system remains a metal. If the LDA bandwidth is much smaller
than the local Coulomb interaction one has essentially
the atomic problem where it costs an energy of about $U$
to add an electron and the system is an insulator.
In between, the Mott-Hubbard metal insulator transition
occurs with V$_2$O$_3$ being on the metallic side
whereas  $ {\rm (V_{0.962}Cr_{0.038})_2O_3} $, which has a
0.1-0.2$\,$eV smaller bandwidth, is on the insulating side.

Interpreting the LDA band structure as a one-particle Hamiltonian
$\hat{H}_{{\rm LDA}}^{0}$ and supplementing it with the local Coulomb
interactions  gives rise to
the multi-band many-body Hamiltonian~\cite{Anisimov91}
\begin{eqnarray}
\hat{H} &\!=\!&\hat{H}_{{\rm LDA}}^{0}\!+\!\!{\ U}\sum_{i\;m}\hat{n}_{i m\uparrow }%
\hat{n}_{i m\downarrow } +\!\!\!\!\!\sum_{i\;m\neq
\tilde{m}\;\sigma \tilde{\sigma}}\!\!\!\!(V\!-\delta _{\sigma
\tilde{\sigma}}J)\;\hat{n}_{i m\sigma
}\hat{n}_{i \tilde{m}\tilde{\sigma}}.
\label{Hint}
\end{eqnarray}%
Here, $i$ denotes the lattice site and $\hat{n}_{i m\sigma }$
is the operator for the occupation of the $t_{2g}$ orbital $m$
with spin $\sigma\in\{\uparrow,\downarrow\}$. The interaction parameters are related by $%
V=U-2J$ which is a consequence of orbital rotational symmetry.
This holds exactly for degenerate orbitals and is a good
approximation in our case where the
$t_{2g}$ bands have similar centers of gravity and bandwidths.
As in the local spin density approximation (LSDA),
the spin-flip term of the exchange interaction is not taken into
account in Eq.~(\ref{Hint}). The consequences of this approximation
for states in the vicinity of the Fermi energy
do not seem to be large as comparative calculations
using the non-crossing approximation within DMFT
show.~\cite{Zoelfl} Furthermore, a pair hopping term
proportional to $J$ is neglected since it requires that one orbital
is entirely empty while another is entirely full which
is a rare situation in the solid state and
 corresponds to highly excited states.
For the Hund's rule coupling $J$ we take the constrained
LDA value $J=0.93\,$eV.~\cite{Solovyev} Unfortunately, such
an {\em ab initio} calculation is not feasible for the Coulomb repulsion $U$
since  $U$ depends sensitively on
screening which leads to uncertainties of about 0.5$\,$eV.~\cite{Nekrasov}
 For our present purposes this
uncertainty is too large since V$_2$O$_3$
is on the verge of a Mott-Hubbard metal-insulator transition,
and, thus, small changes of $U$ have drastic effects.
In particular, due to the small differences
in the LDA band structure it is unlikely that for a $U$ value calculated by
constrained LDA, V$_2$O$_3$ is metallic whereas
$ {\rm (V_{0.962}Cr_{0.038})_2O_3} $ is insulating.
Therefore, we adjust $U$ in such a way as to make sure
that these two systems are metallic and insulating, respectively.
{\em A posteriori}, we will compare the adjusted value with those
calculated by constraint LDA calculations and those extracted  from
the experiment.

So far, we did not specify $\hat{H}_{{\rm LDA}}^{0}$.
In principle, it should contain the valence orbitals,
i.e., the oxygen  $2p$ orbitals and the five vanadium $3d$
orbitals per atom and, maybe, some additional $s$ orbitals.
However, for  V$_2$O$_3$ we are in the fortunate situation
that the three $t_{2g}$ bands at the Fermi energy are
well separated from the other orbitals, see Fig.~\ref{fig:dos_iso}.
Therefore, it is
 possible to restrict ourselves to the three
 $t_{2g}$ bands at the Fermi energy which are made up
of the corresponding atomic vanadium $3d$ orbitals with
some admixtures of oxygen $p$ orbitals.
In the case of three degenerate  $t_{2g}$ orbitals,
which is close to our situation where bandwidths and centers
of gravity of the $a_{1g}$ and the doubly-degenerate $e_{g}^{\pi}$
band are very similar, the $\bf k$-integrated Dyson equation
simplifies to become an integral over the DOS~\cite{Held01}
\begin{eqnarray}
G_m(\omega)&\!=\!&\int {\rm d}\epsilon
\frac{N_{m}^{0}(\epsilon )}{\omega+\mu -\Sigma_m (\omega)-\epsilon}.
\label{Dyson}
\end{eqnarray}
Here $G_m(\omega)$, $\Sigma_m (\omega)$, and $N_{m}^{0}(\epsilon)$~\cite{noteN}
are the Green function, self energy, and LDA density of states, respectively, for the
$t_{2g}$ orbital $m$. In principle, $N_{m}^{0}(\epsilon)$ should
contain a double counting correction, which takes
into account that part of the local Coulomb interaction already
included in the LDA. However, this correction results
in the same effect for all three orbitals and, hence,
only translates into a simple shift of the chemical potential $\mu$.
This makes the issue of how to calculate the double counting correction
irrelevant for the present purposes. The (shifted)  $\mu$ has to be controlled
according to the vanadium valency, i.e., in such a way that
there are two electrons in the three bands at the Fermi energy.

Within DMFT the  $\bf k$-integrated Dyson equation (\ref{Dyson})
has to be solved self-consistently together with a one-site
(mean field) problem which is equivalent to
an Anderson impurity model with hybridization  $\Delta_m(\omega')$
fulfilling~\cite{DMFT_georges}
\begin{equation}
[G_m(\omega)]^{-1} + \Sigma_m (\omega) = \omega + \mu -
\int_{-\infty}^{\infty}\! {\rm d} \omega' \;
\frac{\Delta_m(\omega')}{\omega-\omega'}.
\label{AIM}
\end{equation}
The self-consistent solution of the  Anderson impurity model given by
(\ref{AIM}) together with the Dyson equation (\ref{Dyson})
allows for a realistic investigation of materials with strongly correlated electrons.
At small values of $U$ this procedure typically yields a spectrum  with a central quasiparticle
resonance at the Fermi energy and two incoherent Hubbard side bands, while
at larger values of $U$ the quasiparticle resonance disappears
and a metal-insulator transition occurs.~\cite{DMFT_vollha}
This approach has been successfully applied to
a number of transition metal-oxides,~\cite{LDADMFTTMO,Nekrasov}
transition metals,~\cite{LDADMFTTM} and elemental Pu and Ce.~\cite{LDADMFT4f}
For more details and an introduction to the LDA+DMFT approach
we refer the reader to Ref.~\onlinecite{Held01}.

In the present paper, we solve the multi-band
 Anderson impurity model by  QMC,~\cite{QMC}
where by means of the Trotter discretization and
Hubbard-Stratonovich transformations the interacting
 Anderson impurity model is mapped to a sum of non-interacting
problems, the sum being performed by the Monte-Carlo
technique. We employ a Trotter discretization of
$\Delta \tau=0.25\,$eV$^{-1}$ unless noted otherwise and follow Ref.~\onlinecite{Ulmke}
 for the Fourier transformation between
Matsubara frequencies and imaginary time $\tau$.
To obtain the physically relevant spectral
function  we analytically continue
the Green function from Matsubara frequencies
to real frequencies by means of the
maximum entropy method.~\cite{MEM}
The QMC has the  advantage of being numerically exact while
the main disadvantage is that it is restricted to higher temperatures.
The room temperature calculations of this paper were
computationally very expensive, requiring about $200000$ hours CPU
time on the Hitachi {\sc SR8000-F1} at the Leibnitz Rechenzentrum Munich.
For the implementation of QMC
in the context of LDA+DMFT, including flow diagrams,
 see also Ref.~\onlinecite{Held01}.

\section{LDA+DMFT Spectra}
\label{sect5}\label{spectra}

\begin{figure}[htb]
\includegraphics[clip=true,width=\piclen]{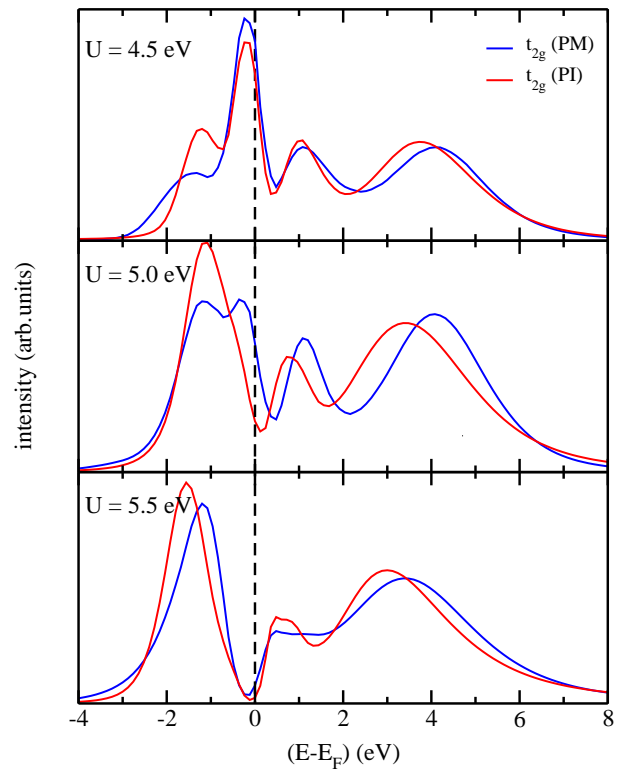}\\
\caption{LDA+DMFT(QMC) spectra for paramagnetic insulating (PI)
${\rm (V_{0.962}Cr_{0.038})_2O_3}$  and metallic (PM) ${\rm V_2O_3}$
at $U=4.5$, $5$, $5.5$ eV, and $T=1160$ K.}
\label{fig:aw_en}
\end{figure}

Using the crystal structure of paramagnetic metallic (PM) ${\rm V_2O_3}$
and paramagnetic insulating (PI) ${\rm (V_{0.962}Cr_{0.038})_2O_3}$,
respectively, as
input, we performed LDA+DMFT(QMC) calculations with one $a_{1g}$
and two degenerate $e_g^\pi$ bands. The results for the spectra of
the $t_{2g}$ bands are shown in Fig.\ \ref{fig:aw_en}. At
$U=4.5$\,eV both crystal structures lead to spectra showing
metallic behavior, with a lower Hubbard band at about $-1$\,eV, an
upper Hubbard band at $4$\,eV and a quasiparticle peak at the Fermi
edge ($0$\,eV). The peak at about $1$\,eV is split from the upper
$t_{2g}$ Hubbard bands due to Hund's rule exchange as we will
discuss below.

By contrast, at $U=5.5$\,eV, both crystal structures lead to
spectra showing nearly insulating behavior. The lower Hubbard band
is strongly enhanced whereas at the Fermi edge, a pseudo-gap is
formed. Above the Fermi energy, only small changes of the two-peak
structure are visible.

Apparently, qualitatively different spectra for the two crystal structures require an
intermediate value of $U$. This is indeed observed at $U=5.0$\,eV:
Whereas pure ${\rm V_2O_3}$ now shows a small peak at the Fermi
edge (a residue of the quasiparticle peak obtained at $U=4.5$\,eV)
and is therefore metallic, the Cr-doped system exhibits a
pronounced minimum in the spectrum implying that it is nearly
insulating. Due to the high temperature of $T=0.1$\,eV $\approx
1160$\,K  of the QMC simulations one only observes a smooth {\em
crossover} between the two phases with a metal-like and
insulator-like behavior of the respective curves instead of a
sharp metal-insulator transition as would be expected for
temperatures below the critical point (i.e., for $T<400$\,K in the
experiment).
 The value of the critical interaction of $5.0$\,eV is in accordance with constrained LDA
calculations by Solovyev {\em et al.}~\cite{Solovyev} who analyzed
the charging energy between di- and trivalent vanadium ions in an
octahedral oxygen environment for LaVO$_3$, obtaining a $U$ value
for the $t_{2g}$ orbitals  which is only slightly smaller than 5
eV. Similar $U$ values of $4-5$\,eV were obtained by fitting
spectroscopy data for vanadium oxides to model calculations.~\cite{U-exp}

\begin{figure}[htb]
\includegraphics[clip=true,width=\piclen]{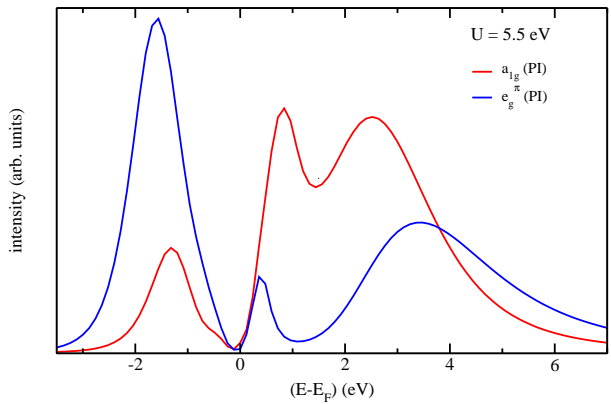}\\
\caption{LDA+DMFT(QMC) spectrum for paramagnetic insulating
${\rm (V_{0.962}Cr_{0.038})_2O_3}$ (PI) for $U=5.5$\,eV; $T=1160$\,K.}
\label{fig:band_res}
\end{figure}

Not only the overall $t_{2g}$ DOS but also the band-resolved
spectra of the $a_{1g}$ and $e_{g}^{\pi}$ bands provide valuable
insight. In Fig.\ \ref{fig:band_res}, the $a_{1g}$ and one of the two
degenerate $e_{g}^{\pi}$ spectra are shown at $U=5.5$\,eV.
The basic features of the spectrum can be understood as
follows: We will show in the next section that the predominant local
configuration has two spin-aligned electrons in the
$e_g^\pi$ orbitals, i.e., a $e_g^\pi e_g^\pi$ spin-1
configuration, with some admixture of  $a_{1g} e_g^\pi$
spin-1 configurations.
Since there are more $e_g^\pi$ than $a_{1g}$ electrons,
let us for a moment disregard the $a_{1g} e_g^\pi$ configurations.
The lower Hubbard band at about $-1.5$\,eV
indicates the removal of an $e_g^\pi$ electron from the
predominantly  $e_g^\pi e_g^\pi$ spin-1 configurations. In an
atomic picture (which is a reasonable starting point for the
insulating phase) this $e_g^\pi e_g^\pi$ $\rightarrow$ $e_g^\pi$
transition leads to an energy gain of $V-J-\mu\approx -1.5$\,eV
(the approximate position of the lower Hubbard band). On the other
hand, the upper $e_g^\pi$ Hubbard band describes the
 $e_g^\pi e_g^\pi$ $\rightarrow$ $e_g^\pi e_g^\pi e_g^\pi$
transitions. Since the spin-alignment is lost this transition
costs an energy $U+V-\mu=U+J+(V-J-\mu)\approx 4.4$ eV, which
roughly agrees with the position of the upper Hubbard band.
On the other hand, adding an $a_{1g}$ electron costs
$2 V-\mu\approx 2.6$\,eV or $2 V-2 J-\mu\approx 0.7$\,eV, depending
on whether this electron is spin aligned or not. This Hund's rule
splitting results in the two-peak structure of the upper $a_{1g}$
Hubbard band in Fig.\ \ref{fig:band_res}.
Since there are also $a_{1g} e_g^\pi$ spin-1 configurations, one has some
modifications of this picture, in particular, there is also a splitting of the upper
$e_g^\pi$ band, resulting in a small peak at $U\approx 0.5$\,eV due to
$a_{1g} e_g^\pi$ $\rightarrow$ $a_{1g} e_g^\pi e_g^\pi$ transitions.
As a consequence of the splitting of the upper Hubbard band into a two-peak
structure, leading to a peak at about $0.5$\,eV above the Fermi edge,
the  gap in the insulating phase is very small, much smaller than
$V\approx 3\;$eV which would be expected in a one-band Hubbard
model. This also explains the puzzle in the attempt to model the
optical gap with a one-band Hubbard model~\cite{Rozenberg95}:
fitting to the small experimental gap one is led to an
unrealistically small Coulomb repulsion of about $1\;$eV and a
bandwidth of less than 0.5$\;$eV.

\begin{figure}[htb]
\includegraphics[clip=true,width=\piclen]{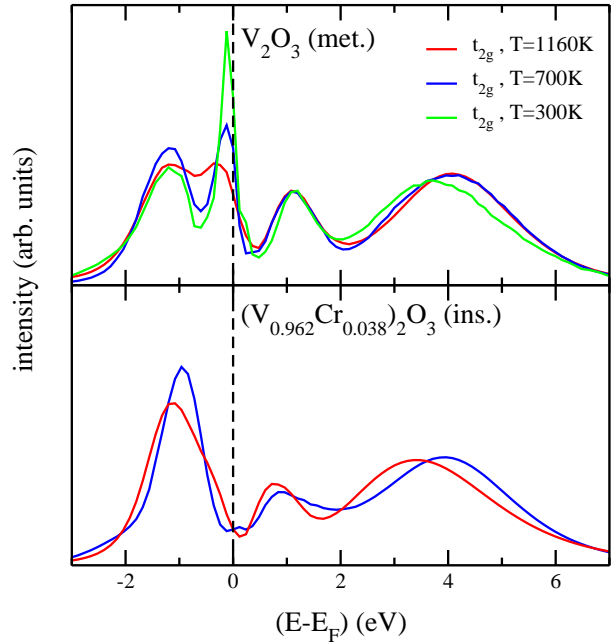}\\
\caption{LDA+DMFT(QMC) spectra for paramagnetic
insulating ${\rm (V_{0.962}Cr_{0.038})_2O_3}$ and metallic ${\rm V_2O_3} $ at
 $U=5$~eV.}
\label{fig:aw_temp}
\end{figure}

To study the metal-insulator transition at experimentally relevant
temperatures, we performed calculations at $T=700$\,K and
$T=300$\,K. Since the computational effort is proportional to
${T^{-3}}$, those low temperature calculations were
computationally very expensive. Fig.\ \ref{fig:aw_temp} shows the
results of our calculations at $T = 1160$\,K, $T=700$\,K, and
$T=300$\,K for metallic ${\rm V_2O_3}$ and for insulating ${\rm
(V_{0.962}Cr_{0.038})_2O_3}$. In the metallic phase, the
incoherent features are hardly affected when the temperature is
changed, whereas the quasiparticle peak becomes sharper and
more pronounced at lower temperatures. This behavior also occurs
in the Anderson impurity model~\cite{Hewson} and has its origin in
the smoothing of the Abrikosov-Suhl quasiparticle resonance at
temperatures  larger than the Kondo temperature. However, in
contrast to the Anderson impurity model this smoothing occurs at
considerably lower temperatures which is apparently an effect of the
DMFT self-consistency cycle.

\begin{figure}[htb]
\includegraphics[clip=true,width=\piclen]{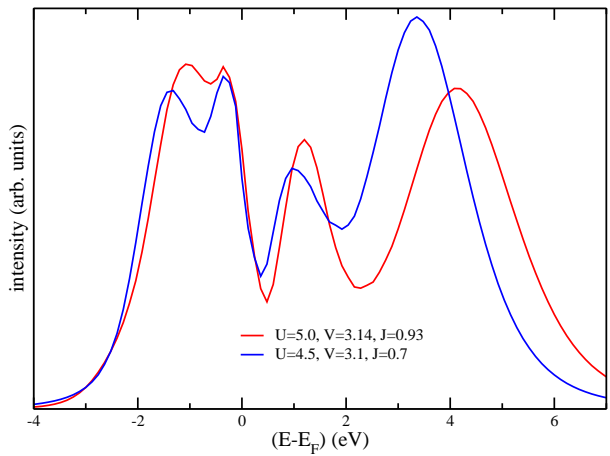}\\
\caption{Comparison of the LDA+DMFT(QMC) ${\rm V_2O_3} $ spectra
at two strengths of the  exchange interaction:
$J=0.93\,$eV (as obtained from constrained LDA), $U=5.0\,$eV, $V=3.14\,$eV,
and $J=0.7\,$eV, $U=4.5\,$eV, $V=3.1\,$eV.}
\label{fig:awhund}
\end{figure}

To study the possible effect of a smaller Hund's rule coupling
$J$, we performed additional calculations for a reduced value of
$J=0.7\,$eV, keeping $V$ nearly constant. The results  in Fig.\
\ref{fig:awhund} show that the positions of the upper Hubbard
bands are significantly shifted towards lower energies while the spectra
below the Fermi energy are hardly affected. This suggests that the
physical properies do not change much. Indeed, we find, for
example, that the spin-1 state hardly changes when the
Hund's exchange is reduced. Even at values as low as
 $J=0.5\,$eV the local moment stays almost maximal, i.e.,
$\langle m_z^2 \rangle=3.85$ at $J=0.5\,$eV, implying that
unrealistically small values of $J$ are required for the
Castellani {\em et al.}~\cite{castellani78}
picture to hold.\cite{foot_castellani}

\section{Changes across the Mott-Hubbard transition}
\label{mit}
\subsection{Local magnetic moment and orbital occupation}

\begin{figure}[htb]
\includegraphics[clip=true,width=\piclen]{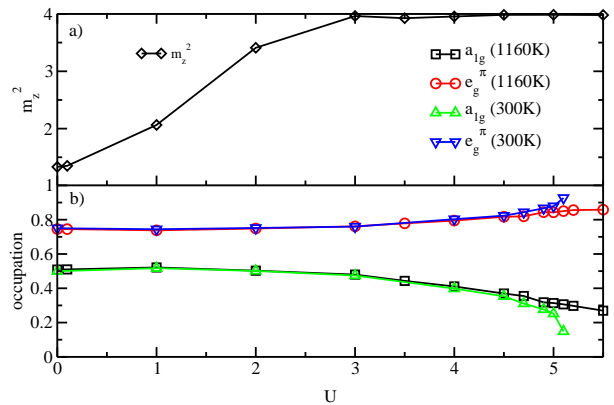}\\
\caption{a) Spin, and b) orbital occupation vs.\  Coulomb
interaction $U$ for metallic $\rm V_2O_3$.}
\label{fig:occupation}
\end{figure}

The spin and orbital degrees of freedom play an important role in
the paramagnetic phase of $\rm V_2O_3$ and in the changes
occurring across the MIT. For example, we find the squared local
magnetic moment $\left<m_{z}^{2}\right>\!=\!\left<\left(\sum_{m}
[\hat{n}_{m\uparrow}-\hat{n}_{m\downarrow}]\right)^2\right>$ to
have a value of $\left<m_{z}^{2}\right>\approx 4$, unaffected
by the MIT, see Fig. \ref{fig:occupation}. This value
corresponds to two  spin-aligned electrons in the ($a_{1g}$,
$e_{g1}^{\pi }$, $e_{g2}^{\pi }$) orbitals and therefore to a
spin-1 state  in the Mott-Hubbard transition regime in good
agreement with polarization dependent x-ray absorption
measurements of Park {\em et al.}~\cite{park} It also agrees with
measurements of the high temperature susceptibility which give the
value of $\mu_{eff}=2.66 \mu_B$ for the effective magnetic moment.~\cite{S-exp}
This is close to the ideal $S=1$ value
$\mu_{eff}=2.83 \mu_B$. Note that when $U$ is reduced to $U < 3$,
the Hund's rule coupling $J$ needs to be reduced as well
to avoid an unphysical attractive Coulomb interaction
(namely, a Coulomb energy $U-3J<0$ would otherwise be gained when
a spin-aligned electron is added to a singly occupied site).
It is this reduction of $J$ which finally leads to a smaller local squared
magnetic moment. Our results of a spin state which is essentially
unaffected by the MIT is in stark contrast to results for the
one-band Hubbard model where $m_{z}^{2}$ changes strongly at the
MIT~\cite{DMFT_georges} (in fact, this quantity had even been used as an indicator for the MIT).

The orbital occupation  (Fig. \ref{fig:occupation}) obtained by us
 clearly rules out a $a_{1g}$ singlet since this would
correspond to $n_{a_{1g}}=1$, $n_{e_{g}^{\pi}}=1$. Therefore
our results contradict the model of Castellani {\em et al.}~\cite{castellani78} who
proposed the formation of an $a_{1g}$ singlet and hence a spin-1/2
state. At all $U$-values we find
predominantly occupied $e_g^\pi$ orbitals, but with a significant
admixture of $a_{1g}$ orbitals (see Fig. \ref{fig:occupation}). On
the basis of an analysis of their linear dichroism data Park {\em
et al.}~\cite{park} concluded that the ratio of the configurations
$(e_g^\pi,e_g^\pi)$ and $(e_g^\pi,a_{1g})$ is equal to 1:1 for the
paramagnetic metallic phase (PM) and 3:2 for the paramagnetic
insulating phase (PI). This corresponds to an electron occupation
of the ($a_{1g}$, $e_{g1}^\pi$, $e_{g2}^\pi$) orbitals of
(0.4,0.8,0.8) for the PI phase and (0.5,0.75,0.75) for the PM
phase. At $T=1160$\,K we find for the PI phase ("insulating"
crystal structure and $U=5.0$\,eV) occupations of (0.28,0.86,0.86),
while for the metallic phase ("metallic" crystal structure and
$U=5.0$\,eV) we obtain (0.37,0.815,0.815). While our results give a
smaller value for the admixture of  $a_{1g}$ orbitals (even more
so at $T=300$\,K), the tendency for the decrease of this value at
the transition to the insulating state is well reproduced. Fig.
\ref{fig:occupation} also shows that in the immediate vicinity of
the Mott transition the orbital occupation has a considerable
temperature dependence, with even fewer electrons in the $a_{1g}$
orbitals at lower temperatures.

Further experimental evidence for a $(e_g^\pi,e_g^\pi)$
configuration in the ground state of the V$^{+3}$ ions in $\rm
V_2O_3$ comes from Brown {\em et al.}~\cite{brown} They measured
the spatial distribution of the field-induced magnetization in
paramagnetic $\rm V_2O_3$ by polarized neutron diffraction. Their
results show that the moment induced on the V atoms is almost
entirely due to the electrons in the doubly degenerate $e_g^\pi$
orbitals with only a minor contribution from the $a_{1g}$ orbital.
For the antiferromagnetic insulating phase of $\rm V_2O_3$,
calculations of the electronic structure by LDA+U
also yielded a spin-1 ground state for the V$^{3+}$ ion and a
$(e_g^\pi,e_g^\pi)$ orbital configuration.~\cite{ezhov99}

The origin for the ground state orbital configuration discussed
above is easily understood from the LDA DOS (Fig.
\ref{fig:dos_met}) where the center of gravity of the $a_{1g}$
orbital is 0.3 eV higher in energy than the corresponding value
for $e_g^\pi$ orbitals. This shift together with the asymmetry of
the DOS leads to an LDA occupation of about 0.55 for the $a_{1g}$
and 0.72 for each of the $e_g^\pi$ bands. The occupation of
$e_g^\pi$ orbitals is further enhanced in the strongly correlated
metallic regime and, in particular, in the insulating phase where
the Coulomb interaction value ($U > 5$\,eV) is significantly larger
than the bandwidth ($W \approx 2$\,eV).

\subsection{Quasiparticle renormalization and spectral weight at the Fermi level}

\begin{figure}[htb]
\includegraphics[clip=true,width=\piclen]{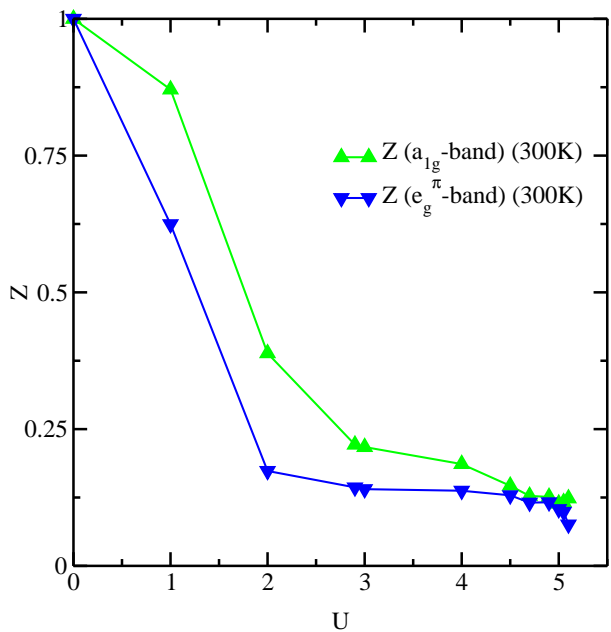}\\
\caption{Quasiparticle weight $Z$ for the $a_{1g}$ and
the $e_g^\pi$ bands vs.\ $U$, using the crystal structure of metallic ${\rm V_2O_3}$.}
\label{fig:z}
\end{figure}

To study the MIT in detail we have calculated the quasiparticle
weight $Z$ by fitting a  third order polynominal to the imaginary part of the QMC
self-energy $\Im \Sigma(i \omega_n$) at the lowest Matsubara frequencies
$\omega_n$ which gives $Z=(1-\partial\Im \Sigma(i \omega)/\partial i
\omega)^{-1}$ via the slope of the polynominal at $\omega=0$. The
resulting quasiparticle weight for the $a_{1g}$- and the $e_g^\pi$-bands
 is shown as a function of $U$ in Fig.\ \ref{fig:z}.
With increasing $U$, $Z$ first shows a strong decrease for both
types of bands. However, in the vicinity of the MIT at about $U =
5$\,eV the $a_{1g}$ quasiparticle weight {\em remains constant}
while that of the $e_g^\pi$ electrons goes to zero. This behavior
of the $a_{1g}$ quasiparticle weight is in striking contrast to
the behavior at the MIT in the one-band Hubbard model where
$Z\rightarrow 0$, such that its inverse, the effective mass,
diverges. Indeed, from Fig.\  \ref{fig:z} alone one might conclude
that the MIT occurs only for the $e_g^\pi$ band.
On the other hand, the total LDA+DMFT spectrum clearly shows
insulating behavior at large Coulomb interactions.

\begin{figure}[htb]
\includegraphics[clip=true,width=\piclen]{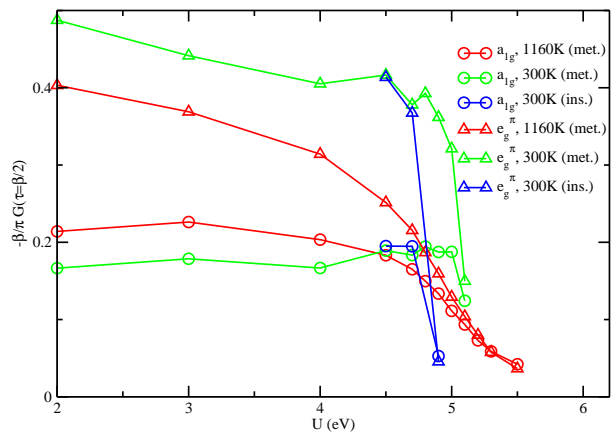}\\
\caption{Spectral weight of the  $a_{1g}$ and $e_g^\pi$ orbitals
at the Fermi energy, as estimated by
 $-\beta/\pi G(\tau=\beta/2)$, vs.\ $U$.}
\label{fig:gtau}
\end{figure}

A quantity which measures the spectral weight
 at the Fermi energy  and does not dependent on the
analytical continuation to real frequencies is given by
\begin{equation}
-\frac{\beta}{\pi} G(\tau=\beta/2) = \int  A(\omega) \underbrace{\frac{\beta}{\pi}\frac{ \exp(-\beta/2 \; \omega)}{1+\exp(-\beta\omega)}}_{K(\omega)} {\rm d} \omega.
\end{equation}
More specifically, $-\beta/\pi G(\tau=\beta/2)$ measures
$A(\omega)$ in the region given by the  kernel $K(\omega)$ which
is centered around the Fermi energy at $\omega=0$ and has a width
proportional to $T=1/\beta$. The results in Fig.\ \ref{fig:gtau}
show that, for $300\,$ K, the spectral weight  at the Fermi energy
disappears at a critical value of $U$ between $5.1$ and $5.2$\,eV
for both types of orbitals in the case of metallic ${\rm V_2O_3}$.
These values of $U$  agree quite well with the
position where the $e_g^\pi$ quasiparticle weight is expected to
disappear in Fig.\ \ref{fig:z}. For the $300\,$ K data for insulating
$ {\rm (V_{0.962}Cr_{0.038})_2O_3} $, the critical U-value is between
$4.9$ and $5.0$\,eV.
With increasing temperature,  the MIT is a smeared out to become a
crossover and, at 1160 K, is only signaled by a change of
curvature slightly above $5\,$eV.

This analysis reaffirms the correctness of our choice of the value
$U=5\,$eV for the simultaneous description of the metallic and
insulating behavior of ${\rm V_2O_3}$ and
$ {\rm (V_{0.962}Cr_{0.038})_2O_3} $, respectively.

\begin{figure}[htb]
\includegraphics[clip=true,width=\piclen]{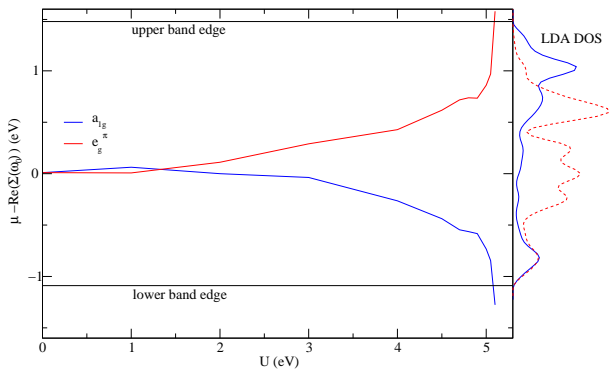}\\
\caption{Effective chemical potential $\mu - \Re \Sigma(\omega_0)$ vs.\ $U$.
The upper and lower band edges of the non-interacting LDA DOS are shown
as solid lines and the entire LDA DOS of $\rm V_2O_3$ is plotted vertically at the
right z-axis.}
\label{fig:mu-re}
\end{figure}

We still have to address the question why the $a_{1g}$
quasiparticle weight remains constant across the transition,
i.e., why the effective $a_{1g}$ mass does {\em not} diverge. The
DMFT Green function is given by Eq. (\ref{Dyson}) which under the
assumption of a Fermi liquid-like self-energy
\begin{equation}
\Sigma_m(\omega) = \Re \Sigma_m(0) + \left.
\frac{\partial \Re \Sigma_m(\omega)}{\partial \omega}\right|_{\omega=0} \omega
\end{equation}
and $Z=(1-\partial \Re\Sigma_m(\omega)/{\partial \omega}|_{\omega=0})^{-1}$ yields
\begin{eqnarray}
G_m(\omega)&\!=\!&\int {\rm d}\epsilon
\frac{Z N_{m}^{0}(\epsilon )}{\omega+Z(\mu -\Re\Sigma_m (0)- \epsilon)}.
\end{eqnarray}
Hence, an MIT can either occur if the effective mass diverges,
i.e., $Z\rightarrow 0$, or if the effective chemical potential
$\mu -\Re\Sigma_m (0)$ moves, due to electronic correlations,
outside the non-interacting LDA DOS such that $N_{m}^{0}(\mu
-\Re\Sigma_m (0))=0$. In the case of V$_2$O$_3$ the latter happens
as is demonstrated by Fig.\ \ref{fig:mu-re}, where $\Re
\Sigma_m(0)$ has been approximated by its value at the lowest
Matsubara frequency, $\Re \Sigma_m(\omega_0)$. At the MIT, between
$U=5.1$ and 5.2$\,$eV, $\mu -\Re \Sigma_{a_{1g}}(0)$ crosses the
upper LDA band edge while  $\mu -\Re \Sigma_{e_g^\pi}(0)$ moves
below the lower band edge.

This explains the pronounced changes of
the orbital occupation in Fig.\ \ref{fig:occupation} and, in
particular, why an MIT can occur although the $a_{1g}$ quasiparticle
weight does not vanish (Fig. \ref{fig:z}). This unexpected feature
of the MIT has important physical consequences: Since at the MIT
$N_{a_{1g}}^0(\mu -\Re \Sigma_{a_{1g}}(0)) \rightarrow 0$, the {\em
height} of the $a_{1g}$ quasiparticle peak goes to zero, rather
than its width which is given by $Z$. For the $e_g^\pi$ band we
have both $N_{e_{g}^{\pi}}^0(\mu -\Re \Sigma_{{e_{g}^{\pi}}}(0)) \rightarrow 0$ {\em
and} $Z\rightarrow 0$ such that {\em height} and {\em width}
simultaneously go to zero. Therefore the quasiparticle DOS
$N_{a_{1g}}^0(\mu -\Re \Sigma_{a_{1g}}(0))/Z$ does not diverge.
Consequently, physical quantities proportional to this
quasiparticle DOS like the linear coefficient of the specific heat
and the local susceptibility do not diverge, at least for the
$a_{1g}$ bands.

\section{Comparison with experimental spectra}
\label{sect6}\label{comparison}

To be able to compare with experimental photo\-emission spectra
the LDA+DMFT results were multiplied with the Fermi function at
the experimental temperature ($T \approx 180$\,K) and broadened
with a $0.09$\,eV Gaussian to account for the experimental
resolution.~\cite{mo02} The same procedure was used for the
comparison with x-ray spectroscopy data (with an inverse Fermi
function at $T=300$\,K and a broadening of $0.2$\,eV taken from experiment). On the
experimental side, the PES of Refs.\ \onlinecite{Schramme00,mo02}
were corrected for the inelastic Shirley-type background which
also removes the O $2p$ contribution.
All experimental and theoretical curves were normalized to yield the
same area (which is a measure of the occupation of the vanadium
$t_{2g}$ bands).

\begin{figure}[htb]
\includegraphics[clip=true,width=\piclen]{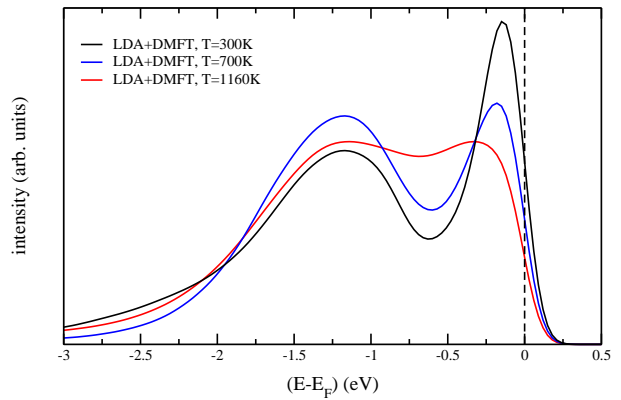}\\
\caption{LDA+DMFT(QMC) results for the metallic phase at different
temperatures.}
\label{fig:Three_theory_curves}
\end{figure}

In Fig. \ref{fig:Three_theory_curves}, LDA+DMFT results for
temperatures $1160$\,K, $700$\,K, and $300$\,K are presented.
Besides the broad, essentially temperature independent peak
at about $-1.25$\,eV corresponding to the lower Hubbard band, the
three curves clearly show the development of a well-defined resonance-like
structure just below the Fermi energy when the temperature is decreased.
The latter peak is what remains of the quasiparticle peak after multiplication with
the Fermi function. At $1160$\,K it is nearly
equal in height to the lower Hubbard band, and there remains
almost no minimum between these two features.
%
% The three curves show essentially the same two features: a broad
% peak at about $-1.25$\,eV corresponding to the lower Hubbard band,
% and a feature near the Fermi energy ($-0.2$\,eV) which is merely a
% shoulder at high temperatures, but develops into a well-pronounced
% separate peak as the temperature is lowered. This second peak is
% what remains of the quasiparticle peak after multiplication with
% the Fermi function. The lower Hubbard band is almost the same for
% the three temperatures, and is in good agreement with the
% photoemission data.\cite{foot_mem}
% With increasing temperature the peak near the Fermi energy
% broadens and decreases in amplitude. At $1160$\,K it is nearly
% equal in height to the lower Hubbard band, and there remains
% almost no minimum between these two features.

\begin{figure}[htb]
\includegraphics[clip=true,width=\piclen]{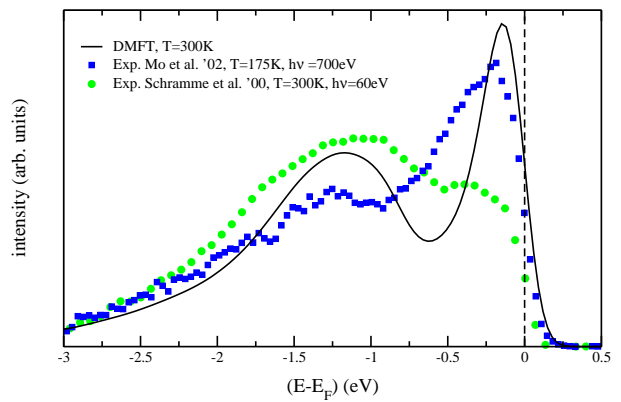}\\
\caption{Comparison of LDA+DMFT(QMC) results at $T=300$\,K with
photoemission data by Schramme {\em et al.}~\cite{Schramme00} and
Mo {\em et al.}~\cite{mo02} for metallic V$_2$O$_3$.} \label{fig:PES_mo}
\end{figure}

In Fig. \ref{fig:PES_mo}, the LDA+DMFT results at $300$\,K are compared with early
photoemission spectra by Schramme~\cite{Schramme00} and recent
high-resolution bulk-sensitive photoemission spectra by Mo {\em et
al.}~\cite{mo02} The strong difference
between the experimental results is now known to be due to the
distinct surface sensitivity of the earlier data. In fact, the
photoemission data by Mo {\em et al.}~\cite{mo02} obtained at
$h\nu = 700$\,eV and $T=175$\,K exhibit, for the first time, a
pronounced quasiparticle peak. This is in good qualitative agreement with
our low temperature calculations. However, the experimental
quasiparticle peak has more spectral weight. The
origin for this discrepancy, for a system as close to a Mott
transition as V$_2$O$_3$, is presently not clear.

\begin{figure}[htb]
\includegraphics[clip=true,width=\piclen]{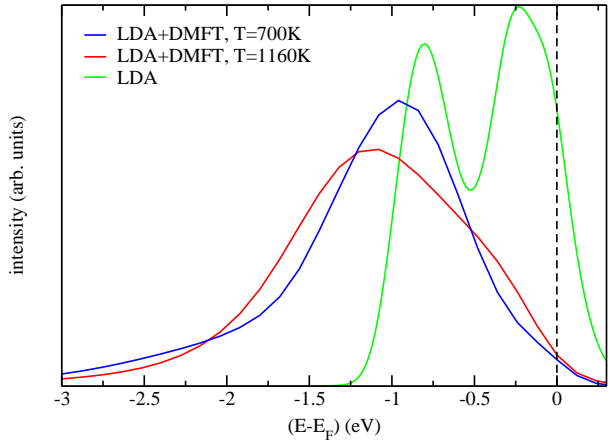}\\
\caption{Comparison of LDA and LDA+DMFT(QMC) results for insulating $ {\rm (V_{0.962}Cr_{0.038})_2O_3} $.}
\label{fig:PES_ins}
\end{figure}

In Fig. \ref{fig:PES_ins} we present the corresponding
calculations for  Cr-doped V$_2$O$_3$: There is a lower Hubbard
band centered at about -1 eV as in the metallic phase, but a quasiparticle
peak at the Fermi energy no longer exists. It is
interesting to note, however, that there remains some spectral
weight in the vicinity of the Fermi energy. Clearly this is not a
Fermi liquid effect, but
is due to highly incoherent states with a large imaginary part of
the low-frequency self-energy.
With decreasing temperatures, this incoherent spectral weight is
reduced and is expected to vanish for $T\rightarrow 0$. Therefore
the resistance {\em increases} with decreasing temperature as is
to be expected for an insulator. For comparison we also show the
LDA data  in Fig. \ref{fig:PES_ins}. They give a completely
different picture: Besides a small peak at about $-0.8$\,eV which
is roughly in the same energy region as the lower Hubbard band of
the LDA+DMFT calculations, it shows a strong peak slightly below
the Fermi energy. Clearly, LDA predicts a metallic solution,
although the input crystal structure is that for insulating
$ {\rm (V_{0.962}Cr_{0.038})_2O_3} $.

\begin{figure}[htb]
\includegraphics[clip=true,width=\piclen]{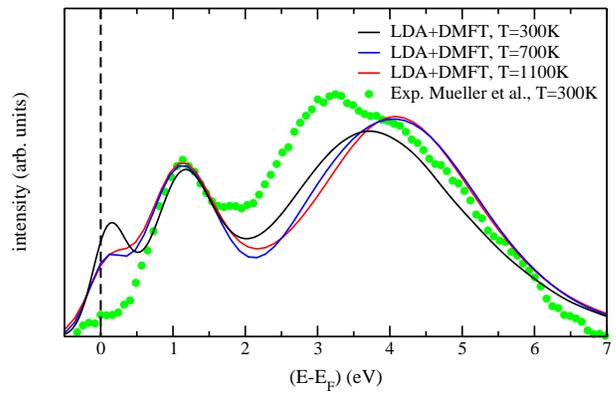}\\
\caption{Comparison of LDA+DMFT(QMC) results with X-ray absorption
data by M\"uller {\em et al.}~\cite{mueller97} for metallic V$_2$O$_3$.}
\label{fig:XAS_mueller}
\end{figure}

While the comparison with PES data provides important insight into
the physics of V$_2$O$_3$, more than half of the
theoretical spectrum lies above $E_F$. For this region we compare our
results at $1160$\,K, $700$\,K, and $300$\,K with O 1$s$ X-ray absorption spectra (XAS) for
V$_2$O$_3$ at $300$\,K by M\"uller {\em et al.}~\cite{mueller97}
(see Fig. \ref{fig:XAS_mueller}). Since in the XAS-data the Fermi
energy is not precisely determined, the data were shifted
so that the peaks at $1.1$\,eV coincide; all
curves were normalized to the same area.

The theoretical spectra above $E_F$ are found to be almost
independent of temperature. Just above the Fermi energy they all
show some structure (i.e., a shoulder at higher temperatures developing
into a small peak at low temperatures ($300$\,K)) which is the residue of
the quasiparticle peak. Furthermore, at $1.1$\,eV there is a rather
narrow peak, and at about $4.2$\,eV a broad peak. The latter two
structures are parts of the upper Hubbard band which is split due
to the Hund's rule coupling $J$. Hence, the relative position of
those two peaks can be expected to depend sensitively on the
value of $J$. A slightly smaller value of $J$ will therefore yield
an even better agreement with experiment.

\begin{figure}[htb]
\includegraphics[clip=true,width=\piclen]{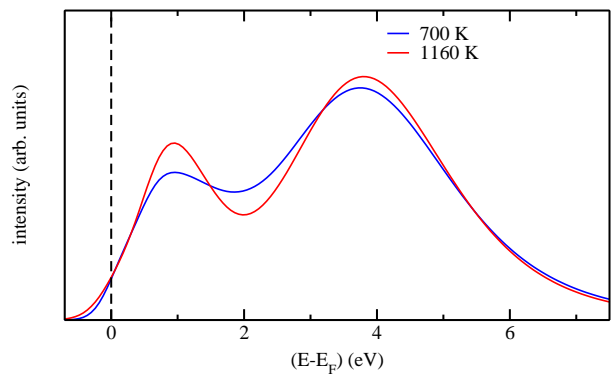}\\
\caption{LDA+DMFT(QMC) spectra for $E>E_F$ for insulating
$ {\rm (V_{0.962}Cr_{0.038})_2O_3} $.} \label{fig:INS_XAS}
\end{figure}

The absence of any quasiparticle weight near $E_F$ in the XAS data
is puzzling. This quasiparticle weight is not only present in the
theoretical spectra above {\em and} below $E_F$, but is also
seen in the high resolution PES measurements by Mo {\em et al.}~\cite{mo02}
below $E_F$. This calls for additional XAS or inverse
photoemission spectroscopy experiments. For comparison with future
experiments, we also show the theoretical XAS spectra for Cr-doped
insulating V$_2$O$_3$ in Fig. \ref{fig:INS_XAS}, where our data
have been broadened  with the experimental resolution of M\"uller
{\em et al.}~\cite{mueller97}

\begin{figure}[htb]
\includegraphics[clip=true,width=\piclen]{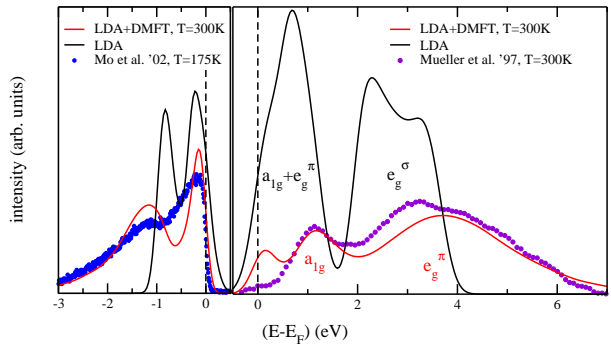}\\
\caption{Comparison of LDA+DMFT(QMC) results with PES data by Mo
{\em et al.}~\cite{mo02} and X-ray absorption data by M\"uller {\em et al.}~\cite{mueller97}
for the metallic phase above and below $E_F$} \label{fig:PES_XAS}
\end{figure}

The comparison between theoretical and experimental spectra for
metallic V$_2$O$_3$ is summed up in Fig. \ref{fig:PES_XAS} where
our LDA+DMFT results for $300$\,K are shown together with the
experimental PES data by Mo {\em et al.}~\cite{mo02} and XAS data by M\"uller
{\em et al.}~\cite{mueller97} To document the theoretical improvement achieved by
including the electronic correlations with the LDA+DMFT technique we also show the results of LDA. We
note again that, by adjusting the value of $U$ such that the
experimentally determined crystal structures lead to the correct
metallic and insulating behavior, the spectrum was calculated
without any further parameter fit. In consideration of this fact
the agreement of our results with the experimental spectra above
{\em and} below the Fermi energy is remarkably good.
Although LDA yields the same gross features, their weight,
position and width neither agree with LDA+DMFT nor experiment. The
interpretation of the two large peaks in the upper half of the
spectrum is also different within LDA and LDA+DMFT. As denoted in
the figure, the peaks in LDA are purely from $t_{2g}$ (lower peak)
and $e_g^\sigma$ (upper peak) bands whereas they are mainly
$a_{1g}$ for the lower and $e_g^\pi$ for the upper peak in
LDA+DMFT, with some admixture of the respective other band.

We note that the $e_g^\sigma$ bands were not taken into account in
our calculations. Therefore, while the complete LDA curve is
normalized to an area of 10 (corresponding to ten d-electrons),
the experimental and LDA+DMFT curves are normalized to an area of
6 (corresponding to the six electrons of the $t_{2g}$ bands). We
may estimate the position of the  $e_g^\sigma$ bands in a LDA+DMFT
calculation as following: Assuming that the intra-$t_{2g}$ Coulomb
interaction $V=U-2J\approx 3\,$eV also applies for the interaction
{\em between} spin-aligned $t_{2g}$- and $e_g^\sigma$-electrons, and
taking the difference between the $e_{g}^{\sigma}$- and
$t_{2g}$-band centers of gravity of roughly $2.5\,$eV into
account, we expect the
 $e_{g}^{\sigma}$-band to be located roughly at $2.5\,$eV$+3\,$eV$=5.5\,$eV above the lower Hubbard band
(-$1.5\,$eV), i.e., at about $4\,$eV. With this estimate  we
expect the (upper) XAS peak at around $4\,$eV in Fig.\
\ref{fig:PES_XAS} to be an admixture  of $e_{g}^{\sigma}$ and
$t_{2g}$ states. More precisely, this upper Hubbard band describes
{\em transitions} from $e_{g}^{\pi}e_{g}^{\pi}$ configurations
with two electrons to the three-electron configurations
$e_{g}^{\pi}e_{g}^{\pi}e_{g}^{\pi}$, $e_{g}^{\pi}e_{g}^{\pi}a_{1g}$, and
$e_{g}^{\pi}e_{g}^{\pi}e_{g}^{\sigma}$ (there is a minor admixture of
$e_{g}^{\pi}a_{1g}$ states).

The properties of paramagnetic V$_2$O$_3$ across the MIT obtained
with LDA+DMFT for a multi-band model are thus found to be remarkably
different from those known from the one-band Hubbard model. Indeed,
the orbital degrees of freedom are seen to play an essential
role: They are not only responsible for the high asymmetry of the
spectra below and above the Fermi energy, but are also
required to explain the reduction of the height of the quasiparticle peak at the Fermi energy
when the MIT is approached in the metallic phase, as well as the smallness
of the insulating gap.

\section{Conclusion}
\label{sect7}\label{summary}

Using LDA-calculated densities of states for paramagnetic metallic
V$_2$O$_3$ as well as paramagnetic insulating
$ {\rm (V_{0.962}Cr_{0.038})_2O_3} $ as input, we performed
DMFT(QMC) calculations at 300\,K, 700\,K, and 1160\,K for various
$U$ values. For $U\approx 5$\,eV, the calculated spectra show a
Mott-Hubbard MIT (or rather a sharp crossover at
the temperatures accessible by present-day QMC simulations).
The details of this MIT are quite different
from those obtained within the one-band Hubbard
model.~\cite{DMFT_georges,DMFTMott_a,DMFTMott_b,DMFTMott_c,Rozenberg95}
In the latter model the {\em height} of the
quasiparticle peak at the Fermi energy is fixed and the MIT is
signaled by a divergence of the effective mass (or the inverse
quasiparticle weight $1/Z$) such that the {\em width} of the
quasiparticle peak goes to zero. In contrast, our LDA+DMFT results
show that, for the $a_{1g}$  quasiparticle peak, the {\em height} goes to zero
while the {\em width}  stays constant, as indicated by a roughly constant
value of $1/Z$ at the MIT. For
the $e_g^\pi$ quasiparticle peak a combination of, both, a reduced
{\em height} and {\em width} at the MIT is found.
This new type of physics, but also the high asymmetry of the spectra below
{\em and} above the Fermi energy as well as the smallness of the
insulating gap, are all due to the {\em orbital degrees of
freedom}.

We compared our theoretical data at $U = 5$\,eV with the results of
various experimental measurements and found the orbital
occupation to be predominantly of $e_g^\pi$ character (with a small admixture of
$a_{1g}$) in agreement with experiments. The occupation decreases
for higher $U$-values, especially at low temperatures. Furthermore,
we found a spin-1 state across the MIT in agreement with
polarization dependent X-ray absorption measurements. The 300 K
spectrum calculated by us for metallic V$_2$O$_3$ is in good
overall agreement with new bulk-sensitive PES measurements.~\cite{mo02}
On the other hand, the difference in the
quasiparticle weight remains to be explained. The comparison with
X-ray absorption measurements shows that our LDA+DMFT(QMC)
calculations also give a good description of the spectrum above
the Fermi energy.

All calculations described above were done
using the integral over the LDA density of states (DOS) (equation
(\ref{Dyson})) to obtain the lattice Green function. For a
non-cubic system, this procedure is an approximation to the exact
LDA+DMFT  scheme. In the future we plan to make use of the full Hamiltonian $H^0$
(eq. (\ref{Hint})).
In this way it will be possible
to study the influence of correlation effects on all
orbitals including the $e_g^\sigma$ orbitals and the oxygen states.

\section{Acknowledgments}
We thank J. W. Allen, Th. Pruschke, M. Feldbacher, I. Nekrasov,  I. S. Elfimov,
A. I. Lichtenstein, A. I. Poteryaev and G. A. Sawatzky for valuable discussions.
This work was supported in part by the Deutsche
Forschungsgemeinschaft through Sonderforschungsbereich 484,
and the Emmy Noether program, by the Russian Foundation for Basic
Research Grant No. RFFI-01-02-17063, and by the
Leibniz-Rechenzentrum, M\"unchen. We thank A. Sandvik for making his maximum entropy
code available to us.


\begin{thebibliography}{99}

\bibitem{Mott} N.\ F.\ Mott, Rev.\ Mod.\ Phys.\ {\bf 40}, 677 (1968); {\sl %
Metal-Insulator Transitions} (Taylor \& Francis, London, 1990);

\bibitem{Gebhard} F.\ Gebhard, {\sl The Mott Metal-Insulator Transition} (Springer,
Berlin, 1997).

\bibitem{RMcWh} T.\ M.\ Rice and D.\ B.\ McWhan, IBM J.\ Res.\ Develop.\ {\bf 14}, 251 (1970).

\bibitem{mcwhan70} D.\ B.\ McWhan and J.\ P.\ Remeika,
Phys.\ Rev.\ B {\bf 2}, 3734 (1970).

\bibitem{mcwhan73b} D.\ B.\ McWhan, A.\ Menth, J.\ P.\ Remeika, W.\ F.\ Brinkman, and
T.\ M.\ Rice, Phys.\ Rev.\ B {\bf 7}, 1920 (1973).

\bibitem{HubbardI} J.\ Hubbard, Proc.\ Roy.\ Soc.\ London {\bf A276}, 238 (1963).

\bibitem{Gutzwiller} M.\ C.\ Gutzwiller, Phys.\ Rev.\ Lett.\ {\bf 10}, 59 (1963).

\bibitem{Kanamori} J.\ Kanamori, Prog.\ Theor.\ Phys.\ {\bf30}, 275 (1963).

\bibitem{HubbardIII} J.\ Hubbard, Proc. Roy. Soc. London {\bf A281}, 401 (1964).

\bibitem{BR} W.\ F.\ Brinkman and T.\ M.\ Rice, Phys.\ Rev.\ B {\bf 2}, 4302 (1970).

% 11 ---------------------------------------------------------------------------

\bibitem{Lie68} E.\ H.\ Lieb and F.\ Y.\ Wu, Phys. Rev. Lett. 20, 1445-1448 (1968).

\bibitem{DMFT_vollha} W.\ Metzner and D.\ Vollhardt, Phys.\ Rev.\ Lett.\ {\bf 62}, 324
(1989).

\bibitem{DMFT_georges} A.\ Georges, G.\ Kotliar, W.\ Krauth, and M.\ J.\ Rozenberg,
Rev.\ Mod.\ Phys.\ {\bf 68}, 13 (1996).

\bibitem{brinkman70}
W.\ F.\ Brinkman and T.\ M.\ Rice, Phys.\ Rev.\ B {\bf 2}, 4302 (1970).

\bibitem{DMFTMott_a} G.\ Moeller, Q.\ Si, G.\ Kotliar, M.\ J.\ Rozenberg,
and D.\ S.\ Fisher, Phys.\ Rev.\ Lett.\ {\bf 74}, 2082 (1995).

\bibitem{DMFTMott_b} M.\ J.\ Rozenberg, R.\ Chitra and G.\ Kotliar,
Phys.\ Rev.\ Lett.\ {\bf 83}, 3498 (1999).

\bibitem{DMFTMott_c} R.\ Bulla, Phys.\ Rev.\ Lett.\ {\bf 83}, 136 (1999).

\bibitem{Rozenberg95} M.\ J.\ Rozenberg, G.\ Kotliar, H.\ Kajueter, G.\ A.\ Thomas,
D.\ H.\ Rapkine, J.\ M.\ Honig, and P.\ Metcalf,  Phys.\ Rev.\ Lett.\ {\bf 75}%
,105 (1995).

\bibitem{Rozenberg97a} M.\ J.\ Rozenberg, Phys.\ Rev.\ B {\bf 55}, R4855 (1997).

\bibitem{Han98a} J.\ E.\ Han, M.\ Jarrell, and D.\ L.\ Cox, Phys. Rev. B {\bf 58}%
, R4199 (1998).

\bibitem{Held98a} K.\ Held and D.\ Vollhardt,
 { Euro.\ Phys.\ J.\ B \bf 5}, 473 (1998).

\bibitem{Limelette03} P.\ Limelette, A.\ Georges, D.\ J\'erome, P.\ Wzietek,
P.\ Metcalf, J.\ M.\ Honig, Science {\bf 302}, 89 (2003).

\bibitem{DMFTMott_d} G.\ Kotliar, Eur.\ J.\ Phys.\ B {\bf 11}, 27 (1999).

\bibitem{DMFTMott_e} G.\ Kotliar, E.\ Lange, M.\ J.\ Rozenberg, Phys.\ Rev.\ Lett.\ {\bf 84}, 5180 (2000).

% 21 ---------------------------------------------------------------------------

\bibitem{held01prl} K.\ Held, G.\ Keller, V.\ Eyert, V.\ I.\ Anisimov, D.\ Vollhardt,
Phys.\ Rev.\ Lett.\ {\bf 86}, 5345 (2001).

\bibitem{Anisimov97} V.\ I.\ Anisimov, A.\ I.\ Poteryaev, M.\ A.\  Korotin, A.\ O.\ Anokhin,
and G.\ Kotliar, J.\ Phys.\ Cond.\ Matter {\bf 9}, 7359 (1997);
A.\ I.\ Lichtenstein and M.\ I.\ Katsnelson, Phys.\ Rev.\ B {\bf 57},
6884 (1998).

\bibitem{Held01} K.\ Held, I.\ A.\ Nekrasov, G.\ Keller, V.\ Eyert, N.\ Bl\"umer, A.K.\ McMahan,
 R.T.\ Scalettar, Th.\ Pruschke, V.I.\ Anisimov und D.\ Vollhardt,
``Realistic investigations of correlated electron systems with LDA+DMFT'',
Psi-k Newsletter \#56, 65 (2003) \\
\verb![!http://psi-k.dl.ac.uk/newsletters/News\_56/Highlight\_56.pdf\verb!]!.

\bibitem{mo02} S.-K.\ Mo, J.\ D.\ Denlinger, H.-D.\ Kim, J.-H.\ Park, J.\ W.\ Allen,
A.\ Sekiyama, A.\ Yamasaki, K.\ Kadono, S.\ Suga, Y.\ Saitoh, T.\ Muro, P.\ Metcalf,
G.\ Keller, K.\ Held, V.\ Eyert, V.\ I.\ Anisimov, D.\ Vollhardt,
Phys.\ Rev.\ Lett.\ {\bf 90}, 186403 (2003).

\bibitem{dernier70a} P.\ D.\ Dernier, J.\ Phys.\ Chem.\ Solids {\bf 31}, 2569 (1970).

\bibitem{mattheiss94} L.\ F.\ Mattheiss, J.\ Phys.: Cond.\ Matt.\ {\bf 6}, 6477 (1994).

\bibitem{hohenberg} P.\ Hohenberg and W.\ Kohn, Phys.\ Rev.\ {\bf 136}, B864 (1964).

\bibitem{kohnsham} W.\ Kohn and L.\ J.\ Sham, Phys.\ Rev.\ {\bf 140}, A1133 (1965).

\bibitem{wkg} A.\ R.\ Williams, J.\ K\"ubler, and C.\ D.\ Gelatt, Jr., Phys.\ Rev.\ B {\bf 19}, 6094 (1979).

\bibitem{revasw} V.\ Eyert, Int.\ J.\ Quantum Chem.\, {\bf 77}, 1007 (2000).

\bibitem{allen} J.\ W.\ Allen, Phys.\ Rev.\ Lett.\ {\bf 36}, 1249 (1976).

% 31 ---------------------------------------------------------------------------

\bibitem{castellani78} C.\ Castellani, C.\ R.\ Natoli, and J.\ Ranninger,
Phys.\ Rev.\ B {\bf 18}, 4945 (1978), ibid. {\bf 18}, 4967 (1978),
ibid. {\bf 18}, 5001 (1978).

\bibitem{park} J.-H.\ Park  {\em et al.},
Phys.\ Rev.\ B {\bf 61}, 11 506 (2000).

\bibitem{ezhov99}
S.\ Yu.\ Ezhov, V.\ I.\ Anisimov, D.\ I.\ Khomskii, and G.\ A.\ Sawatzky,
Phys.\ Rev.\ Lett.\ {\bf 83}, 4136 (1999).

\bibitem{mila} F.\ Mila, R.\ Shiina, F.-C.\ Zhang, A.\ Joshi, M.\ Ma, V.\ Anisimov, T.\ M.\ Rice,
Phys\ Rev.\ Lett.\ {\bf 85}, 1714 (2000).

%\bibitem{tanaka} A.\ Tanaka, J.\ Phys.\ Soc.\ Japan, v.71,N 4,1091-1107, 2002.
\bibitem{tanaka} A.\ Tanaka, J.\ Phys.\ Soc.\ Japan, {\bf 71}, 1091, 2002.

\bibitem{mateo} S.\ Di Matteo, N.\ B.\ Perkins, and C.\ R.\ Natoli,
Phys.\ Rev.\ B \textbf{65}, 054413 (2002).

\bibitem{elfimov} I.\ S.\ Elfimov, T.\ Saha-Dasgupta, M.\ A.\ Korotin, cond-mat/0303404

\bibitem{jellium} L.\ Hedin and B.\ Lundqvist, J.\ Phys.\ C: Solid State Phys.
\textbf{4}, 2064 (1971);
U.\ von Barth and L.\ Hedin, J.\ Phys.\ C: Solid State Phys. \textbf{5}, 1629
(1972).

% 41 ---------------------------------------------------------------------------

\bibitem{jellium2} D.\ M.\ Ceperley and B.\ J.\ Alder, Phys.\ Rev.\ Lett.\ 45, 566 (1980).

\bibitem{bulla_costi} R.\ Bulla, T.\ A.\ Costi, D.\ Vollhardt, Phys.\ Rev.\ B\ {\bf 64}, 045103 (2001).

\bibitem{Anisimov91}  V.\ I.\ Anisimov, J.\ Zaanen, and O.\ K.\ Andersen, Phys.
Rev.\ B {\bf 44}, 943 (1991); V.\ I.\ Anisimov, F.\ Aryasetiawan, and
A.\ I.\ Lichtenstein, J.\ Phys.\ Cond.\ Matter {\bf 9}, 767 (1997).

\bibitem{Zoelfl} M.\ Z\"olfl, Ph.D. thesis, Universit\"at Regensburg, 2001.

\bibitem{Solovyev} I.\ Solovyev, N.\ Hamada, K.\ Terakura,
Phys.\ Rev.\ B {\bf 53}, 7158 (1996).

\bibitem{Nekrasov} I.\ A.\ Nekrasov, K.\ Held, N.\ Bl\"umer, V.\ I.\ Anisimov, and D.\ Vollhardt,
Euro.\ Phys.\ J.\ B {\bf 18}, 55 (2000).

\bibitem{noteN} For the $t_{2g}$ partial DOS's we disregarded the small
$t_{2g}$ contributions to the  oxygen bands (in the range -9 to -4 eV)
and renormalized the  DOS's to unity.
This resembles most closely the DOS's which one would obtain
if one down-folded the LDA bandstructure to
an effective three-band Hamiltonian at the Fermi energy.

\bibitem{LDADMFTTMO}  See, e.g., A.\ Liebsch and A.\ Lichtenstein, Phys.\ Rev.\ Lett.
{\bf 84}, 1591 (2000); I.\ A.\ Nekrasov, Z.\ V.\ Pchelkina, G.\ Keller, Th.\ Pruschke,
K.\ Held, A.\ Krimmel, D.\ Vollhardt, V.\ I.\ Anisimov, Phys.\ Rev.\ B {\bf 67}, 085111 (2003);
 L.\ Craco, M.\ S.\ Laad, and E.\ M\"uller-Hartmann, Phys.\ Rev.\ Lett.\ {\bf 91} 156402 (2003);
A.\ Sekiyama, H.\ Fujiwara, S.\ Imada, S.\ Suga, H.\ Eisaki, S.\ I.\ Uchida, K.\ Takegahara,
H.\ Harima, Y.\ Saitoh, I.\ A.\ Nekrasov, G.\ Keller, D.\ E.\ Kondakov, A.\ V.\ Kozhevnikov,
Th.\ Pruschke, K.\ Held, D.\ Vollhardt, and V.\ I.\ Anishimov, cond-mat/0312429;
E.\ Pavarini, S.\ Biermann, A.\ Poteryaev, A.\ I.\ Lichtenstein, A.\ Georges, O.\ K.\ Andersen, and cond-mat/0309102.

\bibitem{LDADMFTTM}
See, e.g., A.\ I.\ Lichtenstein,  M.\ I.\ Katsnelson, and G.\ Kotliar {\bf 87}, 67205 (2001);
S.\ Biermann, A.\ Dallmeyer, C.\ Carbone, W.\ Eberhardt, C.\ Pampuch,
O.\ Rader, M.\ I.\ Katsnelson, and A.\ I.\ Lichtenstein, cond-mat/0112430.

\bibitem{LDADMFT4f} See, e.g., S.\ Y.\ Savrasov, G.\ Kotliar, and E.\ Abrahams,
Nature {\bf 410}, 793 (2001);
M.\ B.\ Z\"olfl, I.\ A.\ Nekrasov, Th.\ Pruschke, V.\ I.\ Anisimov,
and J.\ Keller, Phys.\ Rev.\ Lett.\ {\bf 87}, 276403 (2001);
K.\ Held, A.\ K.\ McMahan, and R.\ T.\ Scalettar Phys.\ Rev.\ Lett.\ {\bf
87}, 276404 (2001); A.\ K.\ McMahan, K.\ Held,
and R.\ T.\ Scalettar Phys.\ Rev.\ B {\bf 67}, 075108 (2003).

\bibitem{QMC}  J.\ E.\ Hirsch and R.\ M.\ Fye, Phys.\ Rev.\ Lett.\ {\bf 56}, 2521
(1986); M.\ Jarrell, Phys.\ Rev.\ Lett.\ {\bf 69}, 168 (1992); M.\ Rozenberg,
X.\ Y.\ Zhang, and G.\ Kotliar, Phys.\ Rev.\ Lett.\ {\bf 69}, 1236 (1992);
A.\ Georges and W.\ Krauth, Phys.\ Rev.\ Lett.\ {\bf 69}, 1240 (1992);
M.\ Jarrell, in {\em
Numerical Methods for Lattice Quantum Many-Body Problems}, edited by
D.\ Scalapino (Addison Wesley, New-York, 1997).

% 51 ---------------------------------------------------------------------------

\bibitem{Ulmke}
M.\ Ulmke, V.\ Jani\v{s}, and D.\ Vollhardt,  Phys.\ Rev.\ B {\bf 51}, 10411 (1995).

\bibitem{MEM}
M.\ Jarrell and J.\ E.\ Gubernatis,
\newblock { Physics Reports \bf 269}, 133 (1996).

\bibitem{U-exp} A.\ T.\ Mizokawa, A.\ Fujimori, Phys.\ Rev.\ B {\bf 48}, 14150
(1993); J.\ Zaanen, G.\ A.\ Sawatzky, J.\ Solid State Chem.\ {\bf 88},
8 (1990).

\bibitem{Hewson}  A.\ C.\ Hewson {\em The Kondo Problem to Heavy Fermions}
(Cambridge University Press, Cambridge, 1993).

\bibitem{foot_castellani} For us to recover the picture of Castellani {\em et al.}\cite{castellani78}
not only an unrealistically small value of $J$ would be required, but the LDA result would also have
to be different. Namely, a splitting of the $a_{1g}$ orbital into a binding and antibinding
peak of {\em equal} weight would be necessary to obtain an $a_{1g}$ singlet with
occupation $n_{a_{1g}}=1$, leaving an unpaired  spin  $s=1/2$ in the $e_g^{\pi}$ orbitals.

\bibitem{S-exp} D.\ J.\ Arnold, R.\ W.\ Mires, J.\ Chem.\ Phys.\ {\bf 48}, 2231 (1968)

\bibitem{brown} P.\ J.\ Brown, M.\ M.\ R.\ Costa, K.\ R.\ A.\ Ziebeck, J.\ Phys.: Cond.
Matter {\bf 10}, 9581 (1998).

\bibitem{Schramme00} M.\ Schramme, Ph.D. thesis, Universit\"{a}t Augsburg, 2000;
 M.\ Schramme {\em et al.} (unpublished).

%  \bibitem{foot_mem} Differences below $-2$\,eV
% may be due to the maximum entropy method used in our calculations.

\bibitem{mueller97}
O.\ M\"uller, J.-P.\ Urbach, E.\ Goering, T.\ Weber, R.\ Barth,
H.\ Schuler, M.\ Klemm, and S.\ Horn,
Phys.\ Rev.\ B {\bf 56}, 15056 (1997).
\end{thebibliography}
\end{document}